# Self-Regulated Artificial Ant Colonies on Digital Image *Habitats*

Carlos Fernandes[1], Vitorino Ramos[2], Agostinho C. Rosa[1]

[1] LaSEEB-ISR-IST, Technical Univ. of Lisbon (IST),
Av. Rovisco Pais, 1, TN 6.21, 1049-001, Lisbon, PORTUGAL
{cfernandes,acrosa}@laseeb.org
[2] CVRM-IST, Technical Univ. of Lisbon (IST),
Av. Rovisco Pais, 1, 1049-001, Lisbon, PORTUGAL
vitorino.ramos@alfa.ist.utl.pt

*Abstract*. **Artificial life models, swarm intelligent and evolutionary computation algorithms are usually built on fixed size populations. Some studies indicate however that varying the population size can increase the adaptability of these systems and their capability to react to changing environments. In this paper we present an extended model of an artificial ant colony system designed to evolve on digital image habitats. We will show that the present swarm can adapt the size of the population according to the type of image on which it is evolving and reacting faster to changing images, thus converging more rapidly to the new desired regions, regulating the number of his image foraging agents. Finally, we will show evidences that the model can be associated with the Mathematical Morphology *Watershed* algorithm to improve the segmentation of digital grey-scale images.**

## I. Introduction

Self-adaptive multi-agent systems can provide tools for problem solving without centralized control or problem information besides a fitness function. Swarm Intelligence (SI), by means of a population of entities acting together and forming a social collective intelligence over a specific environment or landscape, has led to some computational paradigms like *ACO* (*Ant Colony Optimization*) [3], based on *Ant Colony Systems* [5] (ACS), which proved to be effective on several hard computational problems.

Besides the successful use of bio-inspired ant-like systems over graph problem representations as in ACO, suitable for hard combinatorial optimization problem solving, other authors studied how swarms could develop similar emergent behaviors over 2D grids. Chialvo and Millonas work [4] consists in a simple swarm model (local, memoryless, homogeneous and isotropic) which leads to pheromone trails and networks of ant traffic, without boundary conditions, lattice topology or additional behavioral rules. Like *Ant Colony Systems*, Chialvo's model is based on trail formation by pheromone deposition and evaporation. The pheromone provides the positive feedback of the system, while evaporation works as a negative feedback, causing the whole to emerge self-organized behavior [10].

The approach can be of great importance, since in many fields, the problem at hands could be difficult – if not impossible - to represent by any graph or conceptual network. This is particularly probable for digital image processing, pattern recognition, multimodal function dynamic optimization [8,10], among other possible engineering fields.

Recognizing this starting point shift, Ramos [9] adapted Chialvo's model and some of his bio-inspired behavioral equations to study the evolution of *Ant Colony Systems* on digital image habitats, where the homogeneous flat landscape of the original model was replaced by gray-level images, presented to the swarm in the form of 3D landscapes. The agents evolve on the image itself, by reinforcing pheromone levels around pixels with different gray levels as well as perceptual grouping gestalt features. This procedure causes the emergency of pheromone maps which may be a suitable support for edge detection and image segmentation.

To increase the self-adaptability of the model, we introduced a mechanism to eliminate and create ants along the evolution process. Like in [8], also based on Chialvo's work, we created a search algorithm with varying population size that evolves over dynamical landscapes represented by three-dimensional mathematical functions, where the mechanism must act, on each generation, according to the resources of the habitat and the size of the swarm. In [8], that was achieved by giving more chance of reproduction to ants located on promising regions and neither surrounded by too many ants, nor standing alone on inhabited regions. With this simple method the swarm exhibits self-regulated population size. Also, for each function, that same size converged around a different value, independently of the initial population size. The system, not only showed to be faster than the fixed size population version when led to find higher or lower regions of the landscape, as it also proved to be more reactive to changing landscapes (the change in the environment was simulated by replacing one 3D function by another during evolution process).

With ant colonies evolving in digital images it is not possible to evaluate if a region on which an ant is standing is promising or not, unless a local search mechanism is introduced. In reference [9] model, the ants generally (i.e. stochastically) tend to deposit more pheromone in regions where they perceive a higher difference between the gray levels of the pixels (higher contrast regions tend to acquire more pheromone, while gray level homogenous zones of the image will receive less pheromone). So, it is not possible to know if an ant is reaching optimal regions, unless it is already there. On the other hand, ants tend to reach desired regions via pheromone trail following. With this in mind we conceived a reproduction mechanism based on occupancy rate of the region, as in [8], as well as dependent of the pheromone intensity on the parent's location. This approach resulted on self-regulated swarms that easily adapt to dynamic environments. The results reinforce the idea, expressed in [8], that *Swarms with Varying Population Size* (SVPS) provides a better model to mimic natural features than *Swarms with Fixed Population Size* (SFPS), and improve the population's ability to evolve self-organized foraging behavior, while maintaining a self-regulated population adapted in real-time to different constraints in different search landscapes, or as we will see in the present case, for different images.

The present work is organized as follows; Section 2 gives an overview of related work in the area of artificial life models with varying population size. Section 3 describes the swarm model on which this work is based. The proposed model and its properties are described in section 4. In Section 5 the results are shown and its implications are discussed. Finally, Section 6 concludes the paper and suggests future research.

II. RELATED WORK

Swarm intelligence evolving on digital images was proposed by Ramos and Almeida in [9]. The authors concluded that artificial ant colonies can react and adapt to any type of digital habitat. Since Ramos approach was the support of the swarm proposed in this paper, it will be described in more detail in the next section.

An algorithm for swarm-based color segmentation is presented in [13], where two specialized type of agents evolve over RGB images. The authors applied the algorithm to the segmentation of digital images of butterflies and concluded that solutions to color segmentation problems emerge from the swarm behavior. They also noticed that the solutions identically match segmentations produced by deterministic algorithms.

Reference [14] presents a study on image processing with artificial swarm intelligence. The author's based their algorithm in *Ant Systems* as well and, using perceptual graphs, applied it to the edge-detection of different gray-level images. They conclude that artificial swarms can perform feature extraction in digital images.

Bocchi, Ballerini and Hässler [2] present an evolutionary swarm algorithm for image segmentation where different populations of individuals compete to occupy a bi-dimensional landscape representing the image to be processed. Every cell of the image is occupied by an individual, which has a certain probability to survive in each generation, according to the adaptation of the population to the environmental conditions. If an individual dies, the empty cell is occupied by a newly generated individual. The authors compare their approach to other segmentation algorithms and conclude that the segmentation of noisy images is improved by the swarm algorithm.

On the other hand, Genetic Algorithms (GAs) are usually implemented using populations with fixed size. However, seeking for an extra in performance, in [1], the GAVaPS - Genetic Algorithm with Varying Population Size - was presented, and many works in this area have been made since then. In this specific GA, the concept of "age" of an individual is introduced. A chromosome remains in the population (i.e. stays "alive") for a number of generations proportional to its fitness (*lifetime*). When the age of a chromosome, which is incremented each generation, reaches its lifetime, the individual is removed from the population. Consequently, there is no direct selection pressure in this GA. The individuals are randomly selected for crossover and mutation. The pressure is assured by the fact that fitter individuals remain in the population during a larger number of generations, thus producing more offspring than those with lower fitness. The authors tested the algorithm with four different test functions and concluded that it outperforms the Simple GA in some cases. They also showed that the population size, after a large initial growth, decreases, and remains stable around the initial value. Although the results appeared to be promising, the test functions were not very demanding, and the GAs, simple and with varying population size, attained near-optimal solutions in few function evaluations (between 1000 and 2000). The GAVaPS was tested in [7] with the Royal Road function R4 and in [6] with Spear's multimodal functions. The population behavior described above was not observed in these studies: the population either decreased towards mass extinction, or dramatically increased to enormous dimensions. These results exposed some weak points of GAVaPS, an algorithm that appears to evolve away from a self-organized state and a self-regulated population size. The reproduction mechanism is not constrained by the search space, thus resulting on frequent overpopulation and resulting waste of computational resources. In order to achieve a self-organized behavior and a self-regulated population it is essential to review the reproduction strategy of GAVaPS. The work presented in this paper may also lead to an insight into the proper directions to follow when creating Evolutionary Algorithms with varying population size.

III. THE ANT COLONY ON DIGITAL IMAGES

In the swarm model designed to evolve on digital image *habitats* [9] a certain number of ants is randomly placed on a toroidal bi-dimensional lattice, with the same of the image, which is represented by an *NxN* array, with values between 0 and 255, according to the 8 bit gray level of the pixels. On each generation, each ant moves to an adjacent cell and reinforces the pheromone level on that spot. The model has two different policies: one cell may be occupied by one and only one ant, or ants are allowed to share the same cell. In the first, an ant will not move if it finds itself totally surrounded by others ants. In the previous work [9], this configuration revealed to be more appropriate to evolve on digital images and it was the one used in this study.

| 1/2 | 1 | 1/2 |
|---|---|---|
| 1/4 | Ant | 1/4 |
| 1/12 | 1/20 | 1/12 |

Fig. 1. Values of $w(\Delta\theta)$ used when deciding the next move of an ant coming from south.

The movement of the ants is as in [4]. An ant chooses which cell to move according to its current direction and the pheromone intensity on the eight surrounding cells. That is, if an ant comes from south, and the eight cells have no pheromone, the chance of going north is higher, followed by the chance of going northeast or northwest, an so on, until the likelihood of returning south, which is very low. This is represented by the function $w(\Delta_\theta)$, a probabilistic directional bias, where the domain consists of $\Delta_\theta=0º$, $\Delta_\theta=\pm45º$, $\Delta_\theta=\pm90º$, $\Delta_\theta=\pm135º$ $\Delta_\theta=180º$. As represented in fig. 1 for an ant moving from south, the following values for $w(\Delta\theta)$ were used: $w(0º)=1$, $w(\pm45º)=1/2$, $w(\pm90º)=1/4$ and $w(180º)=1/20$ (discrete landscapes were considered, so $\Delta_\theta=\pm45$, for instance, means the northeastern or southwestern pixel from the current location, for an ant moving from south).

The final probabilities, however, are dependent on the pheromone level found on the neighboring cells. The relative probabilities to move to a cite $i$ with pheromone density $\sigma_i$ are given by (1), where $W(\sigma)$ is given by (2). Since a two dimensional lattice is used, $i$ represents the eight cells that surround $k$.

$$P_{ik} = \frac{W(\sigma_i)w(\Delta_\theta)}{\sum_{j/k}W(\sigma_j)w(\Delta_\theta)} \quad (1)$$

In (2), $\beta$ represents the osmotropotaxic sensitivity (controls the tendency to follow pheromone: high values of $\beta$ result in swarms heavily attracted by pheromone, while lower values cause the swarm to behave in a more randomly way) and $1/\delta$ is the sensory capacity, which describes the fact that each ant's ability to sense pheromone decreases somewhat at high concentrations.

$$W(\sigma) = \left(1 + \frac{\sigma}{1+\delta\sigma}\right)^\beta \quad (2)$$

The probabilities $P_{ik}$ depend on two weighting factors: $w(\Delta_\theta)$ reflects the directional bias, while $W(\sigma)$ is proportional to the pheromone already deposited on the inspected cell. After probability's computation, the direction is chosen by a roulette-well based method.

Every time an individual visits a cell it adds a constant amount $\eta$ of pheromone, plus a dynamic value $p\Delta_{gl}$ (equation 3, where $p$ is a constant and $\Delta_{gl}$ represents the difference between the median grey-levels of previous cell and its neighbors, and current cell and its neighbors). In addition, in every iteration all cell's pheromone is decreased by a constant amount $K$.

$$T = \eta + p\Delta_{gl}/255 \quad (3)$$

With this simple mechanism of pheromone deposition/evaporation and its ability to attract ants, an emergent and autocatalytic mass behavior is achieved, leading to significant increase in pheromone around the edges of the elements of the image and near highly contrasted regions. Fig. 2 states a brief description of the swarm model.

IV. SELF-REGULATED VARIATION OF POPULATION SIZE

Reference [8] describes a SVPS based on an ant colony similar to the one described in the previous section, and designed to evolve on 3D landscapes ($N^2 \to N$ functions). The variation of the population size is achieved by an aging (and dying) process as well as a reproduction process. The first was built on the concept of age: the older the ant, the higher will be the chance to die. The reproduction process depends on the occupation of parent's adjacent cells and the height of the parents on the landscape ($f(x,y)$, where $x$ and $y$ represent the position of the ant). For the reasons stressed out in section 1, this approach is, at least, incomplete if we try to apply it to ant colonies on digital images. However, we started from the rules defined in [8] and tried to adapt them to the current problem.

A fixed amount of energy $e(0)$ is assigned to each ant created. Every generation, the energy $e(a)$ is updated, where $a$ is the age of the ant measured in time steps. As it can be seen in (4), $e(a)$ is computed by subtracting a fixed amount to the energy of the ant in the previous iteration, and by adding a dynamic value inspired in the pheromone deposition T of (3). The last reflects the contrast of the region where the ant is moving. So, a high value of $\Delta_{gl}$ means that the ant is making a transition from regions of the image with different gray-level pixels. The term max$\Delta_{gl}$ represents the maximum $\Delta_{gl}$ found so far. This means that the energy of ants moving between two regions where the difference between the gray-level median values is equal to max$\Delta_{gl}$ it will remain the same, since it increases in an amount equal to the fixed amount decreased.

The energy $e(0)$ is set to $1+\alpha$ to assure that all ants make at least one move before they die. Then, at each generation, all the ants' $e(a)$ is computed. Finally, an ant dies with probability $P = 1-e(a)$.

The reproduction cycle inspects all ants' position. An individual (main parent) triggers a reproduction procedure if it finds at least another ant occupying one of its 8 surrounding cells (*Moore* neighborhood is adopted). Then, the probability to

```
/*High-level description of the model*/
For all agents do place agent at randomly at selected cell
End For
For t = 1 to t_max do  /* Main loop */
  For all agents do
    Compute W(σ) and P_ik /* According to equations (1) and (2) */
    Move to a selected adjacent cell not occupied by other ant
    Increase pheromone at cell c
        P(c)= P(c)+[η+pΔ_gl/255]
  End For
  Evaporate pheromone by K, at all cells
End For
```

Fig 2. High-level description of the ant colony model [9].

$$e(t) = e(t-1) - \alpha + \alpha * (\Delta_{gl} / \max \Delta_{gl}) \qquad (4)$$

generate offspring is computed according to equation 5, where *n* is the number of occupied neighboring cell and *W(n)* is a weighting factor depending on that occupancy rate. In the tests described bellow *W(n)* is as [4]: *W(0) = W(8) =0; W(4) = 1; W(5) = W(3) =0.75; W(6) = W(2) =0.5; W*(7) = W(1) = 0.25*. These values tend to favor crossover between ants standing in moderately populated regions of the lattice.

$$P_r = W(n) * (\mu + \frac{(1-\mu)\Delta_{gl}}{\max \Delta_{gl}}) \qquad (5)$$

Notice that the second term in the product is very similar to the pheromone deposition rate *T*. The constant μ assures that even ants moving in homogeneous regions may have some chance to reproduce (if μ is set to 0, ants moving between regions with $\Delta_{gl}=0$ won't have a chance to reproduce).

## V. EXPERIMENTAL SETUP AND RESULTS

To examine the differences between SFPS and SVPS when evolving on static environments, 10 test images - shown in fig. 3 - were used. The first five were taken from [4]. *Baboon*, *F16*, *Lena* and *Peppers* are classical image processing testbeds. Image *IST* was chosen for the purpose of dynamical landscape experiences.

Like in [9], the parameters were set to the following values: β = 3.5; σ = 0.2; η = 0.07; *k* = 1.0; *p* = 1.5; *S*=30% (*S* is the size of the initial population measured in percentage of the total environment space in pixels). The SVPS parameters α and μ were set to 0.025 and 0.1 respectively. These values seem to be appropriate for all the images. They are not image dependent and in SVPS they can be changed with less effect in the quality of results than those observed in SFPS. The size of *Cross*, *Einstein*, *Map*, *Marble*, *Road* and *IST* images is 100x100, while *Baboon*, *F16*, *Lena* an *Peppers* have 200x200 pixels.

Both swarms evolved over the images during 500 time steps. Resulting pheromone distribution maps of SFPS and SVPS can be seen in fig. 4 and fig. 5 respectively. Mass behavior of SVPS appears to be faster in edge detection by pheromone deposition. In fewer generations, the maps of SVPS show a configuration that resembles the main lines of the original image figure. Moreover, for some images, like *Cross* and *Marble*, the pheromone maps have different characteristics: the homogenous regions have less noise in SVPS maps, giving the

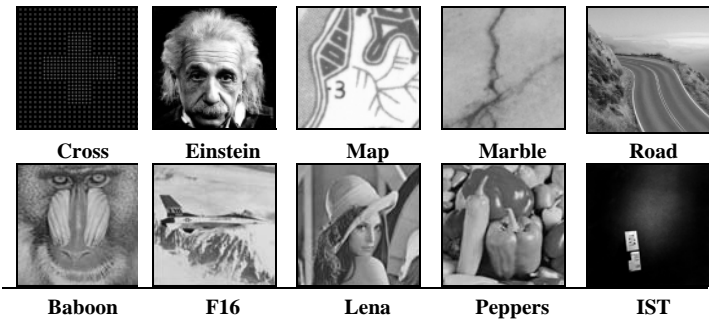

Fig. 3. The ten test images used to analyze the present proposed algorithm.

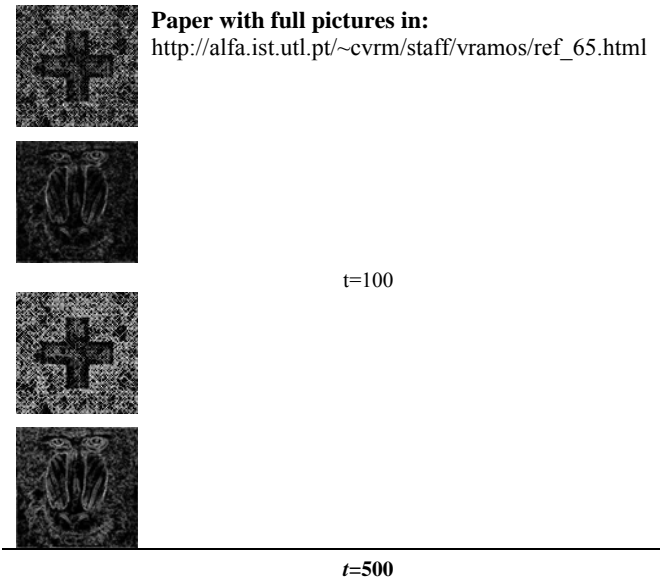

**Paper with full pictures in:**
http://alfa.ist.utl.pt/~cvrm/staff/vramos/ref_65.html

t=100

*t*=500

Fig.4. Pheromone maps created by SFPS (Swarm with Fixed Population size) after 100 and 500 iterations.

idea that ants are converging more efficiently to heterogeneous parts of the image. Also, most of the maps of SVPS at *t*=20 resemble the ones attained with SFPS at *t*=500, regarding noise and sharpness of the pheromone around the edges of the original figures. Further tests at the end of this section will

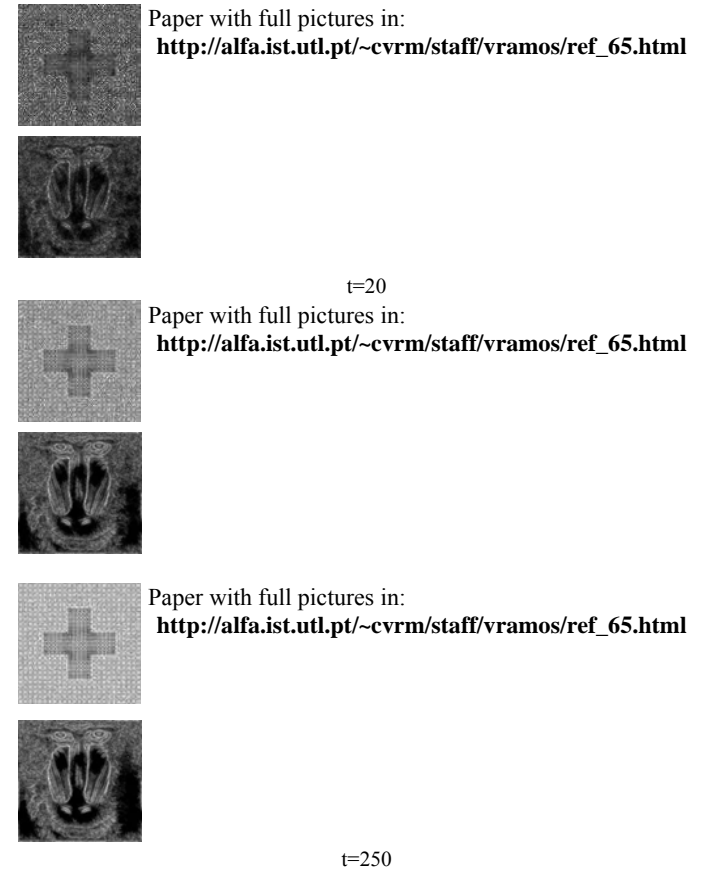

Paper with full pictures in:
**http://alfa.ist.utl.pt/~cvrm/staff/vramos/ref_65.html**

t=20

Paper with full pictures in:
**http://alfa.ist.utl.pt/~cvrm/staff/vramos/ref_65.html**

Paper with full pictures in:
**http://alfa.ist.utl.pt/~cvrm/staff/vramos/ref_65.html**

t=250

Fig. 5. Pheromone maps created by SVPS (Swarm with Variable Population size) after 20, 100 and 250 iterations.

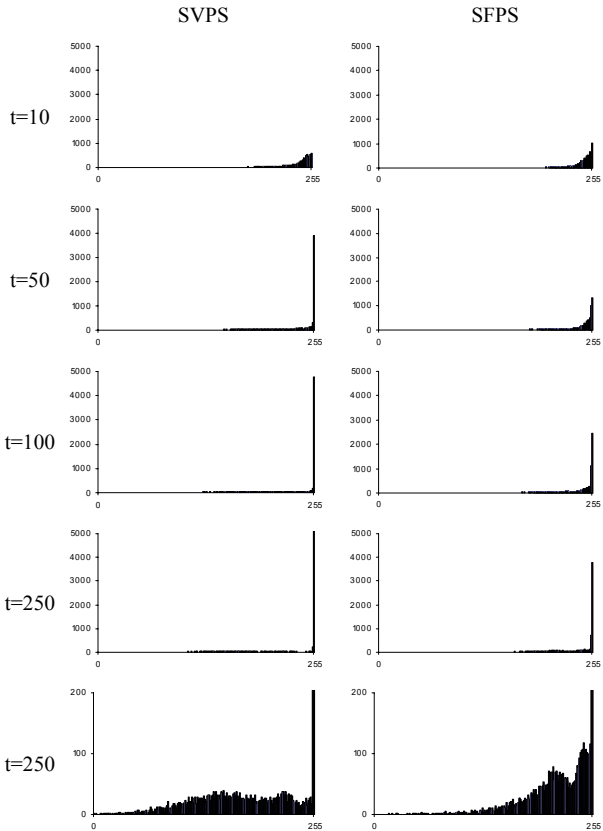

Fig. 6. Gray-level histograms of pheromone maps obtained by SVPS and SFPS with image *Marble*.

demonstrate the convenience of the noiseless pheromone maps achieved by SVPS.

The idea that SVPS creates pheromone maps with less noise and higher intensity regions of pheromone is reinforced by the gray-level histograms shown in fig. 6. By observing the graphs it can be concluded that SVPS on *Marble* emerge a vaster pheromone free region, represented in the histograms by the high column in level 255 (value that represents black pixels in the maps of fig. 4 and fig. 5). By direct observation of the maps in figures 4 and 5 it is clear that SVPS achieves a larger black region without the noise generated from "off the tracks" pheromone deposition. The last row in fig. 6 shows a detail of the histogram at *t*=250. It is clear that SVPS creates a sharper and more contrasted pheromone field while the map of SFPS is still dominated by a large amount of darker gray-level pixels with values near 255. The swarms evolving on other images show the same behavior.

Fig. 4 and fig. 5 display the pheromone maps that emerged from SVPS and SFPS at certain time steps. However, since we are comparing search procedures with fixed and varying population size it may be convenient to analyze the growth of the population of SVPS, since larger populations have higher computational costs. Fig. 7 shows the growth of SVPS population size during 100 time steps. When running over some images, the SVPS population tends to grow larger than 30% of the space. For this reason, figures 4 and 5 compare the pheromone maps of SFPS at *t*=500 and SVPS at *t*=250. Except for the image *Cross*, over which SVPS evolves populations near 8000 individuals, SVPS needs equal or less computational effort to achieve the results shown in both figures – check table 1 for the mean population size of SVPS over 100 iterations on the complete test bed. Remember that SFPS population size equals 30 percent of the search space.

Table 1. Mean population size (measured in percentage of the search space) of SVPS over 100 time steps.

| Cross | Einstein | Map | Marble | Road |
|---|---|---|---|---|
| 80,11 | 47,60 | 62,16 | 33,10 | 47,19 |
| Baboon | F16 | Lena | Pepper | IST |
| 55,47 | 32,03 | 37,86 | 41,41 | 9,8 |

Fig. 7 also shows that the model adapts its population size to the image where it is evolving, suggesting a self-regulating intrinsic mechanism. Although these may seem to be the optimal population values for each image, changing the population size of SFPS in some images to values near the median population size of SVPS does not improve its performance, as we can see in fig. 8 where maps are clearly noisier than those represented in fig. 3. These results suggest that the two swarms have different ways to achieve emergent and autocatalytic behavior, although both are based in the same simple stigmergic [10] rules. The differences observed may be a result of the self-regulated variation of the population size.

Fig. 9 represents the growth of SVPS population with image *Einstein* with starting populations of 1000, 2000, 3000 and 8000 ants. It is clear that the size of the populations converge to similar values after a few hundred generations independently of the initial population size. This outcome was also observed with other images and reinforces the idea that the swarm self-regulates the population size around an optimal value that depends on the *habitat* (digital image).

The second part of the test aims to check the ability of both swarms to evolve in changing landscapes. For that purpose, the SVPS and SFPS worked over one image during a certain number of time steps. Then, the image is replaced by another and the swarm is required to evolve from a pheromone field that reflects previous image characteristics and must readapt itself to the new environment conditions.

Figures 10 and 11 shows the resulting maps of several experiences (the parameters were set as in previous tests, except for experience D where S=60%). In A and B, both swarms (SVPS and SFPS) started with *Einstein* image. At *t*=100 *Einstein* was replaced by *Map*. The aspect of resulting

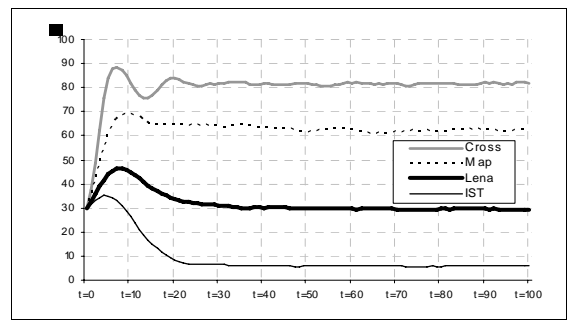

Fig. 7. SVPS population size over 100 generation on images *Cross*, *Map*, *Lena* and *IST*.

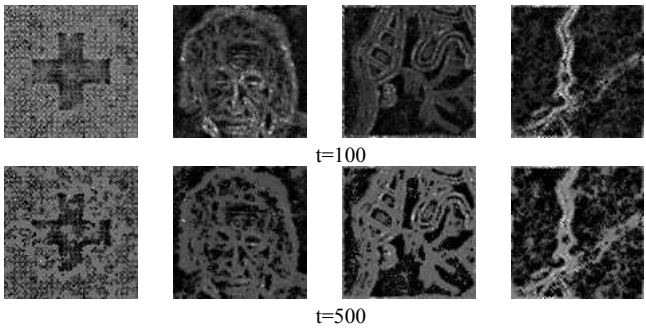

t=100

t=500

Fig. 8. Pheromone fields of SFPS after 100 and 500 generations. The parameters are equal as those in Fig. 2, except *S* which is equal to 80%, 50%, 60% and 40% respectively in *Cross*, *Einstein*, *Map* and *Marble*.

pheromone maps indicate that SVPS readapts the pheromone distribution quickly to the new environment, while SFPS experiences great difficulties to escape from the memory of previous image. Since SVPS population size grows larger than 30% of the search space the SFPS is allowed to run for 1000 iterations. In C, a SFPS was used to evolve on *Einstein* image until *t*=100; then *Einstein* image was replaced by *Map* and SFPS by SVPS. The results appear to be similar to B. In D the SFPS performed the same task as in A but with 6000 ants instead of 3000. Again, the increase in the population size does not seem to bring any benefits.

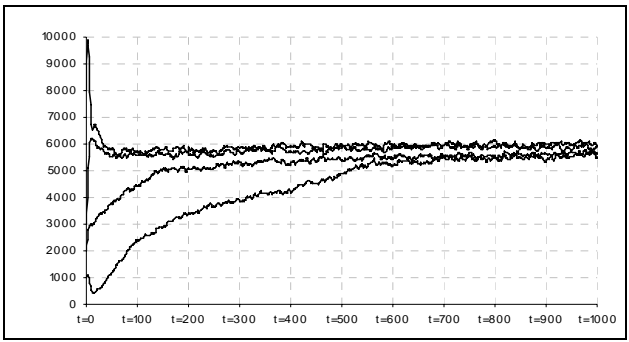

Fig. 9. SVPS population size growth on *Einstein* image for different initial size (1000, 2000, 3000 and 8000 ants).

In experiences E (SFPS) and F (SVPS) the *Einstein* image is replaced at t=100 by *Marble* image. In G and H the swarms start on *Marble* and continue on *Map* after *t*=100. The results clearly show that SVPS is more able to react to changing landscapes by escaping the system memory and evolving new and different pheromone maps adapted to the new image.

SVPS also adapts the size of the population according to the image in which it is evolving at the moment. Fig. 12 shows the population growth - measured in percentage of the search space (image dimension) - in experiences, B, F and H. At *t*=100, when image 1 is replaced by image 2, there is a disturbance in population size which by then was stabilized around a value determined by the landscape. When the image is replaced the swarm tends to adjust the population size to the characteristics of the new image.

One last test was conducted to examine the response of the systems when one image with small and concentrated heterogeneous areas is replaced by another with the same characteristics but with the desired region in a rather different location on the lattice. The swarms started to evolve in image *IST*. At *t*=250 *IST* is replaced by *IST* with a 180º rotation. The results are shown in fig. 13. The ants of SVPS converge massively to the new region after *t*=300 while the population of SFPS is still divided between the new heterogeneous region and the previous location of that same region after 1500 time steps.

By inspecting more closely what happens when images change it can be concluded that the dynamics of SVPS is radically different from the one observed in SFPS population.

A        B        C        D

Paper with full pictures in:
**http://alfa.ist.utl.pt/~cvrm/staff/vramos/ref_65.html**

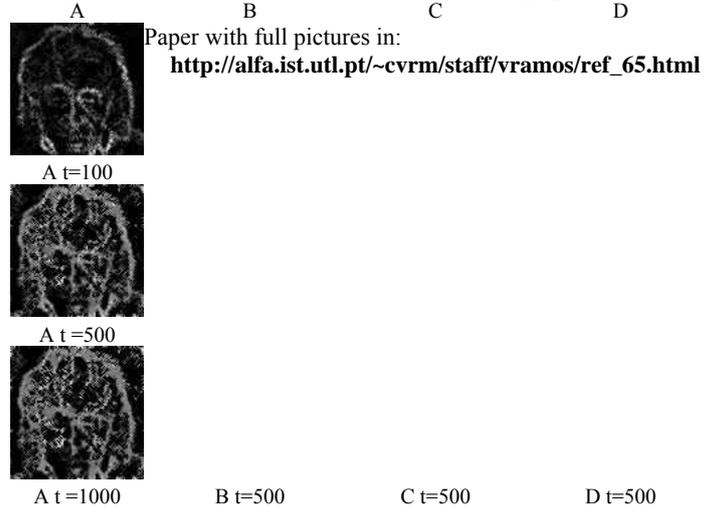

A t=100

A t=500

A t=1000    B t=500    C t=500    D t=500

Fig. 10. Emerging pheromone maps in dynamic landscapes. The swarms start to evolve over *Einstein* image. After 100 iterations (50 in column D) the image changes to *Map*. (A – SFPS; B – SVPS; C – SFPS until *t=100*, then SVPS; D – SFPS). Parameters as figures 3 and 4, except D, where *S=60%*).

E        F        G        H

Paper with full pictures in:
**http://alfa.ist.utl.pt/~cvrm/staff/vramos/ref_65.html**

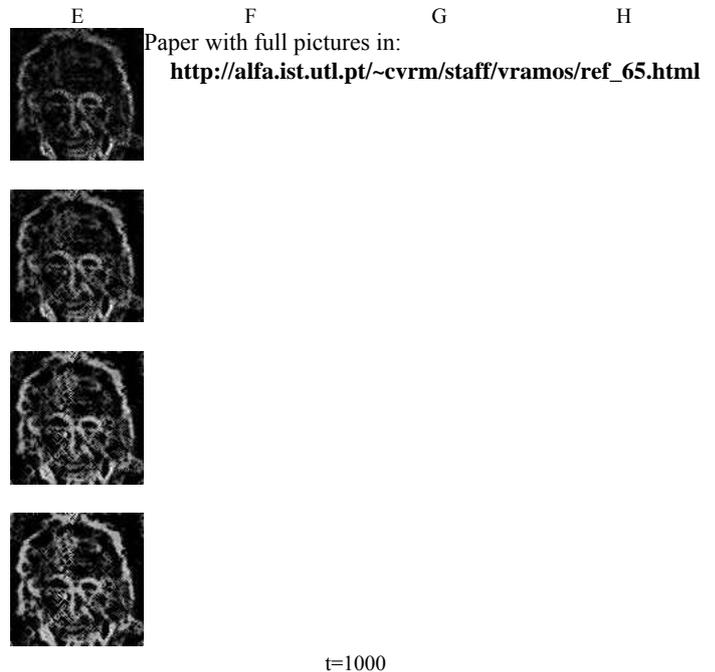

t=1000

Fig. 11. Emerging pheromone maps in dynamic landscapes. In E and F, the swarms start on *Einstein* image which is replaced by *Marble* at *t*=100. In G and H, the first image is *Marble* and the second (after *t*=100) is Map. E and G are the result of SFPS, while experiments F and H where made with SVPS.

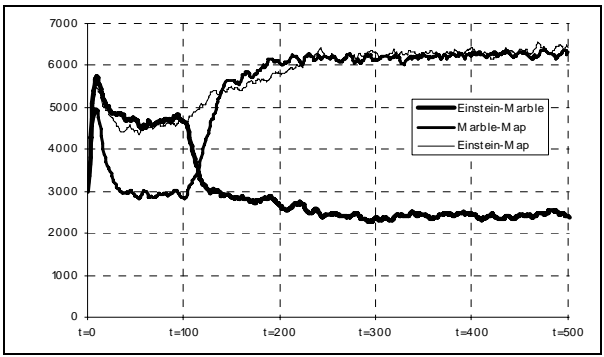

Fig. 12. SVPS population growth in experiences B, E and G.

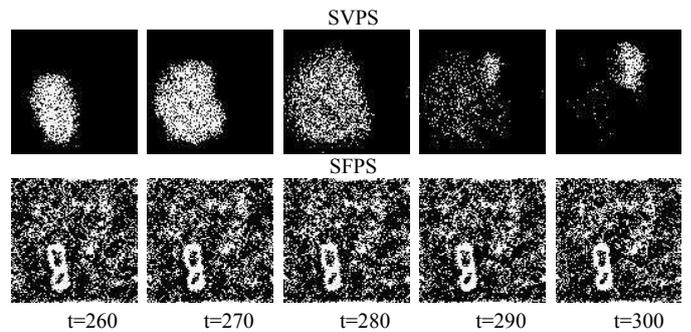

Fig. 14. Ants position in the landscape after *IST* is replaced by rotated *IST*.

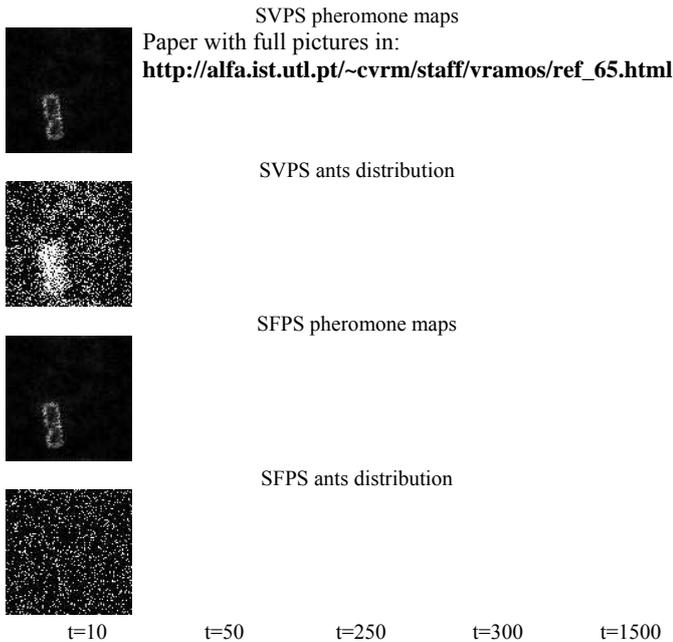

Fig. 13. Pheromone maps and ants distribution resulting from SVPS and SFPS. The image *IST* is replaced at *t*=250 by *IST* rotated by 180º.

In fig. 14 we can see that SVPS population behaves as a flock when the rewarding region is transferred to the opposite side of the landscape, while SFPS gradually reaches that region by a more kind of explorative foraging behavior (rather then exploitive) inherent to the swarm mechanism. SVPS can be seen as an artificial life model where heterogeneous regions work as food resources, making a coherent compromise between exploitation of old regions and exploration of new ones. When those resources are finished the swarm starts a quasi-random search for promising regions. If that region is not close enough, the population eventually experiences an extinction process. These phenomena may be simulated with SVPS by increasing the value of $\beta$ (thus increasing the tendency to follow pheromone) in the previous experience. This way the randomness of the system is decreased and when the image changes and the region that favors reproduction moves to the opposite side of the landscape, the swarm is more reluctant to leave the pheromone high intensity part of the environment causing a dramatically decrease in the mean reproduction probability, leading to the extinction of the popu-

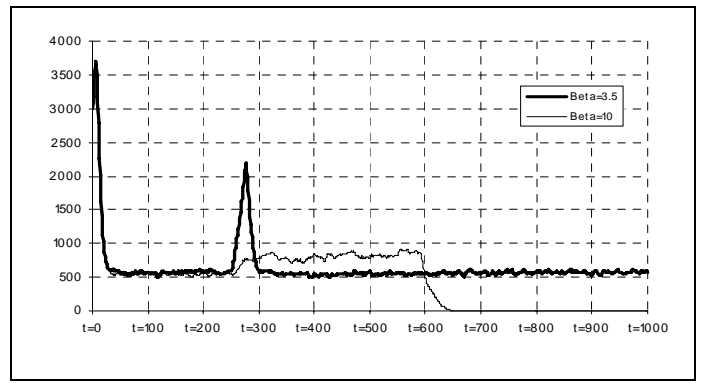

Fig. 15. SVPS population when image *IST* is replaced by image *IST* rotated at *t*=250.

-lation as we can see in fig. 15. Another phenomena worth noting is the disturbance in population size at *t*=250 when image 1 is replaced by image 2. Notice that SVPS with $\beta$=3.5 experiences a sudden increase in population size. This behavior reinforces the idea that the swarm is spreading trough the landscape searching for contrasted regions. Remember that the only pressure towards contrasted pheromone free regions is achieved by the reproduction process. So, the widening of the space occupied by the population is not a consequence of pheromone following but an indirect outcome of the reproduction process. When $\beta$ is increased the population tends to continue exploring the already marked regions and demographic explosion does not occur, as we can see in fig. 15. Without converging massively to heterogeneous regions, where ants have better chances to reproduce, the population eventually decreases to extinction.

## VI. CONCLUSIONS AND FUTURE WORK

It has been shown that SVPS is more able to evolve sharp and noiseless pheromone maps that somewhat reflect the image contours. By self-regulating population size, the model avoids overcrowded swarms which, like in SFPS, can lead to pheromone deposition on homogeneous regions. Also, allowing the SFPS to evolve for more generations (until t=1000, for instance) is not enough to attain pheromone maps as sharp as those attained by SVPS. The varying population model is not only faster, but it is also more effective in creating pheromone trails around the edges of the images. The computational cost of the reproduction process is not significant and the running times SFPS/SVPS are very similar.

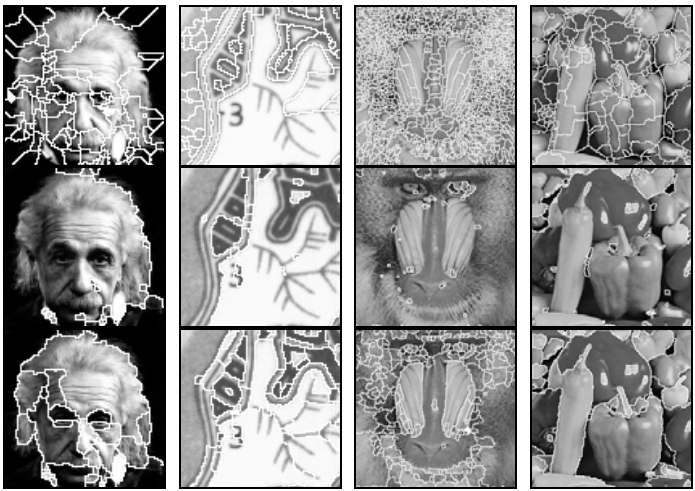

Fig. 16. Segmentation of images *Einstein*, *Map*, *Baboon* and *Peppers* with classical Watershed algorithm [12, 11] (1st row) and a Watershed algorithm based on SFPS (2nd row) and SVPS (last row) pheromone maps.

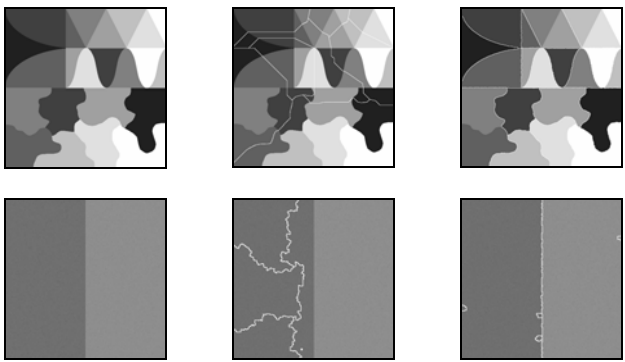

Fig. 17. Segmentation of images *texmos3* and *nstp100* with classical Watershed algorithm [12, 11] (2nd column) and a Watershed algorithm based on SVPS (3rd column) pheromone maps. Original images in (left column).

Some authors [13, 14] point out that swarm intelligence may produce adequate templates to improve image segmentation and edge detection of digital images. In a current study, the pheromone maps obtained by the models have been used as input to the Watershed algorithm [12, 11] to improve the segmentation of gray-level images. Preliminary results are shown in fig.16. In B and C the pheromone maps of SFPS and SVPS were introduced in place of the complement of enhanced image, thus eliminating the previous processing required before going through the last stage of the classical Watershed algorithm [12, 11]. This procedure eliminates the dependence of the structuring elements, since that parameter is used to compute the image which is replaced in the process. In the examples shown, it is clear that SVPS maps lead to more appropriate image segmentation than SFPS. The Watershed algorithm with SVPS maps also appears to attain better results than traditional Watershed as it can be seen in fig. 16 comparing images A and C. (The Watershed segmentation images in A are the best results obtained after several tests with different structuring elements.) Fig. 17 shows segmentation results of Watershed and Watershed plus SVPS pheromone maps for two classical image processing testbeds which are hard to solve by traditional image segmentation algorithms. As it can be seen in the results shown in fig. 17, Watershed algorithm with SVPS pheromone maps appears to overcome the difficulties posed by those images, while Watershed alone performs poorly. However, these results are still preliminary and further investigation is required in order to draw conclusions about the effectiveness of a Watershed algorithm with pheromone maps.

ACKNOWLEDGMENT: The first author wishes to thank FCT, Ministério da Ciência e da Tecnologia, his research Fellowship SFRH/BD/18868/2004.